\documentclass[aps,prl,twocolumn,superscriptaddress,showpacs]{revtex4}

\usepackage{graphicx}
\usepackage{dcolumn}
\usepackage{bm}
\usepackage{ulem}
\usepackage{color}

\newcommand{\al}{$\alpha$}

\newcommand{\raa}{($\alpha$,$\alpha$)}
\newcommand{\raX}{($\alpha$,$X$)}
\newcommand{\raapr}{($\alpha$,$\alpha'$)}
\newcommand{\rag}{($\alpha$,$\gamma$)}
\newcommand{\ran}{($\alpha$,n)}
\newcommand{\raan}{($\alpha$,$\alpha$n)}
\newcommand{\raaa}{($\alpha$,$2\alpha$)}

\newcommand{\raap}{($\alpha$,$\alpha$p)}
\newcommand{\rapp}{($\alpha$,2p)}
\newcommand{\rap}{($\alpha$,p)}

\newcommand{\rga}{($\gamma$,$\alpha$)}

\newcommand{\stot}{$\sigma_{\rm{reac}}$}
\newcommand{\sred}{$\sigma_{\rm{red}}$}
\newcommand{\ered}{$E_{\rm{red}}$}

\newcommand{\zniv}{$^{64}$Zn}

\newcommand{\gevii}{$^{67}$Ge}
\newcommand{\geviii}{$^{68}$Ge}
\newcommand{\gavii}{$^{67}$Ga}

\begin{document}

\title{
Relation between total cross sections from 
elastic scattering and $\alpha$-induced 
reactions: the example of $^{64}$Zn
}

\author{Gy. Gy\"urky}
\email{gyurky@atomki.mta.hu}
\affiliation{
Institute of Nuclear Research (ATOMKI), H-4001 Debrecen, Hungary}
\author{P. Mohr}
\affiliation{
Institute of Nuclear Research (ATOMKI), H-4001 Debrecen, Hungary}
\affiliation{
Diakonie-Klinikum, D-74523 Schw\"abisch Hall, Germany}

\author{Zs. F\"ul\"op}
\affiliation{
Institute of Nuclear Research (ATOMKI), H-4001 Debrecen, Hungary}

\author{Z. Hal\'asz}
\affiliation{
Institute of Nuclear Research (ATOMKI), H-4001 Debrecen, Hungary}

\author{G.G. Kiss}
\affiliation{
Institute of Nuclear Research (ATOMKI), H-4001 Debrecen, Hungary}

\author{T. Sz\"ucs}
\affiliation{
Institute of Nuclear Research (ATOMKI), H-4001 Debrecen, Hungary}

\author{E. Somorjai}
\affiliation{
Institute of Nuclear Research (ATOMKI), H-4001 Debrecen, Hungary}
\date{\today}

\begin{abstract}

The total reaction cross section is related to the elastic scattering angular
distribution by a basic quantum-mechanical relation. We present new
experimental data for $\alpha$-induced reaction cross sections on $^{64}$Zn
which allow for the first time the experimental verification of this simple
relation at low energies
by comparison of the new experimental reaction data to the result
obtained from $^{64}$Zn($\alpha$,$\alpha$)$^{64}$Zn elastic scattering. 

\end{abstract}

\maketitle

A main application of quantum mechanics is nuclear physics. Here basic
theoretical relations can be tested experimentally with high precision. 
An interesting example is the simple relation between the total (non-elastic)
reaction cross section \stot\ and the elastic scattering cross section:
\begin{equation}
\sigma_{\rm{reac}} =
   \frac{\pi}{k^2} \sum_L (2L+1) \, (1 - \eta_L^2) \quad \quad .
\label{eq:stot}
\end{equation}
Here $k = \sqrt{2 \mu E_{\rm{c.m.}}}/\hbar$ is the wave number,
$E_{\rm{c.m.}}$ is the energy in the center-of-mass (c.m.) system, and
$\eta_L$ and $\delta_L$ are the real reflexion coefficients and scattering
phase shifts which define the angular distribution
$\left(\frac{d\sigma}{d\Omega}\right)(\vartheta)$ of elastic
scattering. Eq.~(\ref{eq:stot}) is derived from a partial wave analysis 
using the standard two-body Schr\"odinger equation. 
This Eq.~(\ref{eq:stot}) is widely used, in
particular in the calculation of reaction cross sections using the statistical
model (StM; to avoid confusion with the widely used abb.\ ``SM'' for ``shell
model''). In the following discussion we will focus on \al -induced
reactions at low energies. 

In the StM the reaction cross section of an \al -induced  \raX\ reaction is
calculated in two steps. In the first step the total reaction cross section
\stot\ is calculated using Eq.~(\ref{eq:stot}); the reflexion coefficients
$\eta_L$ are determined by solving the Schr\"odinger equation using a global
\al -nucleus potential, e.g.\ the widely used potential by McFadden and
Satchler \cite{McF66}. Compound formation is the dominating absorption
mechanism at energies from a few MeV up to about several tens of MeV; thus, it
is assumed that the compound formation cross section is approximately given by
the total reaction cross section: $\sigma_{\rm{compound}} \approx$ \stot . In
the second step $\sigma_{\rm{compound}}$ is
distributed among all open channels. The decay branching is obtained from the
transmission factors into the various open channels which are again calculated
using global potentials for each (outgoing particle + residual nucleus)
channel or from the photon strength function in the case of the
\rag\ channel. Further details of the StM can be found in the recent review
\cite{Rau11}. 

To our knowledge, the underlying basic relation in Eq.~(\ref{eq:stot}) has
never been verified experimentally for \al -induced reactions at low energies
around or below the Coulomb barrier. 
Although there is no special reason to suspect to Eq.~(\ref{eq:stot}) for the
particular case of \al -induced reactions, an experimental verification
assures its application for the calculation of reaction cross sections which
has turned out to be difficult especially at low energies (see discussion
below). 
Alternatively, if the validity of Eq.~(\ref{eq:stot}) is assumed {\it{a
    priori}}, it can be used as a stringent test for consistency between
different methods for the determination of \stot .

\stot\ has been measured at higher energies up to 200\,MeV
using transmission experiments \cite{Igo63,Lab73,Auce94,Ing00}. The
experimental results have significant uncertainties, and earlier results
\cite{Auce94} have been questioned later by the same group \cite{Ing00}. 
For the early experiments
\cite{Auce94} it was found that \stot\ from the transmission data is
significantly smaller than \stot\ derived 
from \raa\ elastic scattering using Eq.~(\ref{eq:stot})
\cite{Abe94,Auce94,Ait95} whereas agreement was found using the latest
transmission data \cite{Ing00} where \stot\ is about $10-30$\,\% higher
compared to earlier results \cite{Auce94}. 

At lower energies \al -induced reactions have been studied 
intensively in the last decades
to determine a global \al -nucleus potential which is then used for the
prediction of \al -induced reaction cross sections and their inverse, mainly
\rga , reaction cross sections. Often the motivation came from astrophysics
where \rga\ reaction rates under stellar conditions play an important role for
the nucleosynthesis of heavy neutron-deficient nuclei (the so-called
$p$-nuclei) \cite{Woo78,Arn03,Rau06,Rap06}. It turned out over the years that it
is very difficult or even impossible to obtain a consistent description of
elastic \raa\ scattering and \raX\ cross sections. Especially the reproduction
of \rag\ capture cross sections for heavy targets (above $A \approx 100$) at
the lowest experimentally accessible energies (i.e.\ the most relevant energy
range for the calculation of stellar reaction rates) was poor
\cite{Som98,Gyu06,Ozk07,Cat08,Yal09,Gyu10,Kis11}; with significant efforts
better results have been obtained mainly for \ran\ reactions at
slightly higher energies very recently \cite{Avr10,Sau11,Mohr11,Pal12}.

It is the aim of the present work to provide an experimental confirmation of
Eq.~(\ref{eq:stot}) at relatively low energies around the Coulomb barrier
because it is implicitly used in all the above mentioned studies of \rag\ and
\ran\ reactions
\cite{Som98,Gyu06,Ozk07,Cat08,Yal09,Gyu10,Kis11,Avr10,Sau11,Mohr11,Pal12}. The
target nucleus \zniv\ is an almost perfect candidate for such a study because
($i$) a series of experiments have measured angular distributions of elastic
\raa\ scattering in the energy range from 12 to 50\,MeV 
\cite{DiP04,OLDSCAT} 
which enables to
study the \zniv -\al\ potential in a wide energy range, 
and ($ii$)
almost all relevant reaction channels at low energies lead to unstable
residual nuclei 
which can be measured using the activation technique. Thus, this experiment
can use a completely different experimental approach for the determination of
\stot\ where the systematic uncertainties of transmission experiments can be
avoided (see discussion in \cite{Ing00}).
In addition we point out that the
activation technique is the most appropriate tool for the determination of
total \raX\ reaction cross sections whereas e.g.\ in-beam \rag\ experiments
might miss weak $\gamma$-ray branches. 

Although several data sets for \raX\ cross sections on $^{64}$Zn 
are already available
(e.g.\ \cite{Por59,Rud69,Ste64,Cog65,Scu09}), 
the data quality is relatively
poor. So we have remeasured the cross sections of the \zniv \rag \geviii,
\zniv \ran \gevii , and \zniv \rap \gavii\ reactions at low
energies. Relatively low energies were chosen not only because of the
underlying astrophysical request for a low-energy \al -nucleus potential, but
also because of the relatively small numbers of open channels. In the energy
range under study, the total reaction cross section is given by the following
sum:
\begin{eqnarray}
\sigma_{\rm{reac}} & = &
  \sigma(\alpha,\gamma) +
  \sigma(\alpha,n) +
  \sigma(\alpha,p) +   
  \sigma(\alpha,\alpha') +
  \nonumber \\
& &
  \sigma(\alpha,2\alpha) +   
  \sigma(\alpha,\alpha p) +
  \sigma(\alpha,2p) +
  \sigma(\alpha,\alpha n)
\label{eq:sum}
\end{eqnarray}

In the following we present first our new experimental data for the \zniv \rag
\geviii , \zniv \ran \gevii, and \zniv \rap \gavii\ reactions. Then we
estimate the cross sections of the remaining open channels in
Eq.~(\ref{eq:sum}) which are much smaller than the dominant \rap\ and \ran\
cross sections. Next we compare the sum of the \al -induced cross sections to
the total reaction cross section \stot\ from the analysis of 
an angular distribution of \zniv \raa \zniv\ elastic scattering and find
agreement within the uncertainties, i.e.\ we confirm the basic
quantum-mechanical relation in Eq.~(\ref{eq:stot}). Finally, we suggest
potential improvements to reduce the uncertainties.

The cross sections of the three studied $\alpha$-induced reactions have been
measured with the activation method. The experimental technique was similar to
the one described in one of our recent works \cite{hal12}. Some important
aspects are briefly described here. For further details see also \cite{gyu12}.

The targets were prepared by evaporating natural isotopic composition metallic Zn onto 2\,$\mu$m thick Al foils. The target thicknesses (typically between 100 and 500\,$\mu$g/cm$^2$) have been measured by weighing and Rutherford backscattering spectrometry. The $\alpha$-irradiations have been carried out at the cyclotron accelerator of ATOMKI which provided an $\alpha$-beam with up to 1\,$\mu$A intensity. The durations of the irradiations varied between half an hour and one day. Changes in the beam intensity were taken into account by recording the current integrator counts in multichannel scaling mode with 1 minute time constant. The stability of the targets was continuously monitored by detecting the elastically scattered alpha particles from the target with an ion implanted Si detector built into the chamber at 150 degrees. With a beam intensity not higher than 1\,$\mu$A no target deterioration has been observed.

The cross sections of the \zniv\rag\geviii , \zniv\rap\gavii , and \zniv\ran\gevii\ reactions have been determined from the measurement of the $\gamma$-radiation following the $\beta$-decay of the reaction products having half-lives of 270.93\,d, 3.26\,d, and 18.9\,min, respectively. The decay parameters of the reaction products are summarized in Table\,\ref{tab:decay}. Owing to the largely different half-lives of the reaction products, the gamma-counting of one target has been done two or three times.

\begin{table}
\caption{\label{tab:decay} Decay parameters of the reaction products. Only the strongest gamma transitions used for the analysis are listed. Data are taken from \cite{NDS68} and \cite{NDS67}. Since the decay of \geviii\ is not followed by gamma radiation, the decay of its short-lived daughter, $^{68}$Ga has been measured.
}
\begin{tabular}{cccrr@{$\pm$}l}
\hline
Reaction & Product & half-life 
& \multicolumn{1}{c}{E$_\gamma$} & \multicolumn{2}{c}{relative} \\
         & isotope & 					 
& \multicolumn{1}{c}{[keV]}	& \multicolumn{2}{c}{intensity [\%]} \\
\hline
\zniv\rag & \geviii & 270.93\,d & --- & \multicolumn{2}{c}{---} \\
\geviii\ decay & $^{68}$Ga & 67.71\,min & 1077.3 & 3.22 & 0.03\\
\zniv\rap & \gavii & 3.26\,d & 184.6 & 21.41 & 0.01\\
                             &&& 209.0 & 2.46 & 0.01 \\
                             &&& 300.2 & 16.64 & 0.12 \\
                             &&& 393.5 & 4.56 &0.24 \\
\zniv\ran & \gevii\ & 18.9\,min & 167.0 & 84.28 & 4.52 \\
                             &&&  828.3 & 2.99 & 0.27 \\
                             &&&  1472.8 & 4.9 & 0.2 \\
\hline
\end{tabular}
\end{table}

The $\gamma$-measurements were carried out with a 100\,\% relative efficiency HPGe detector equipped with a 4$\pi$ low background shielding. The absolute efficiency of the detector at a large source-to-detector distance of 27\,cm was measured with several calibrated sources \cite{hal12}. In this geometry the true coincidence summing effect is completely negligible. The strong activity of the short lived \gevii\ reaction product was measured in this far geometry. 

At higher alpha energies the \rap\ cross section is large enough so that the target activity could be measured in far  geometry. At low energies, on the other hand, a close source-to-detector distance of 1\,cm was used where the true coincidence summing effect is strong. In order to take this into account, spectra of strong \gavii\ sources were measured both in far and close geometries and for all studied transitions a conversion factor between the two geometries was calculated. This factor accounts for the ratio of the efficiencies as well as the effect of summing.

The weak source activities from the \zniv\rag\geviii\ reaction could only be measured in the close counting geometry. Owing to the simple decay scheme of $^{68}$Ga, however, the summing effect is negligible here. Therefore, for the  $^{68}$Ga activity measurement the absolute efficiency of the detector has been measured directly in close geometry using summing-free, single line calibration sources ($^7$Be, $^{54}$Mn, $^{65}$Cu and $^{137}$Cs). 

\begin{figure}
\includegraphics[angle=270,width=\columnwidth]{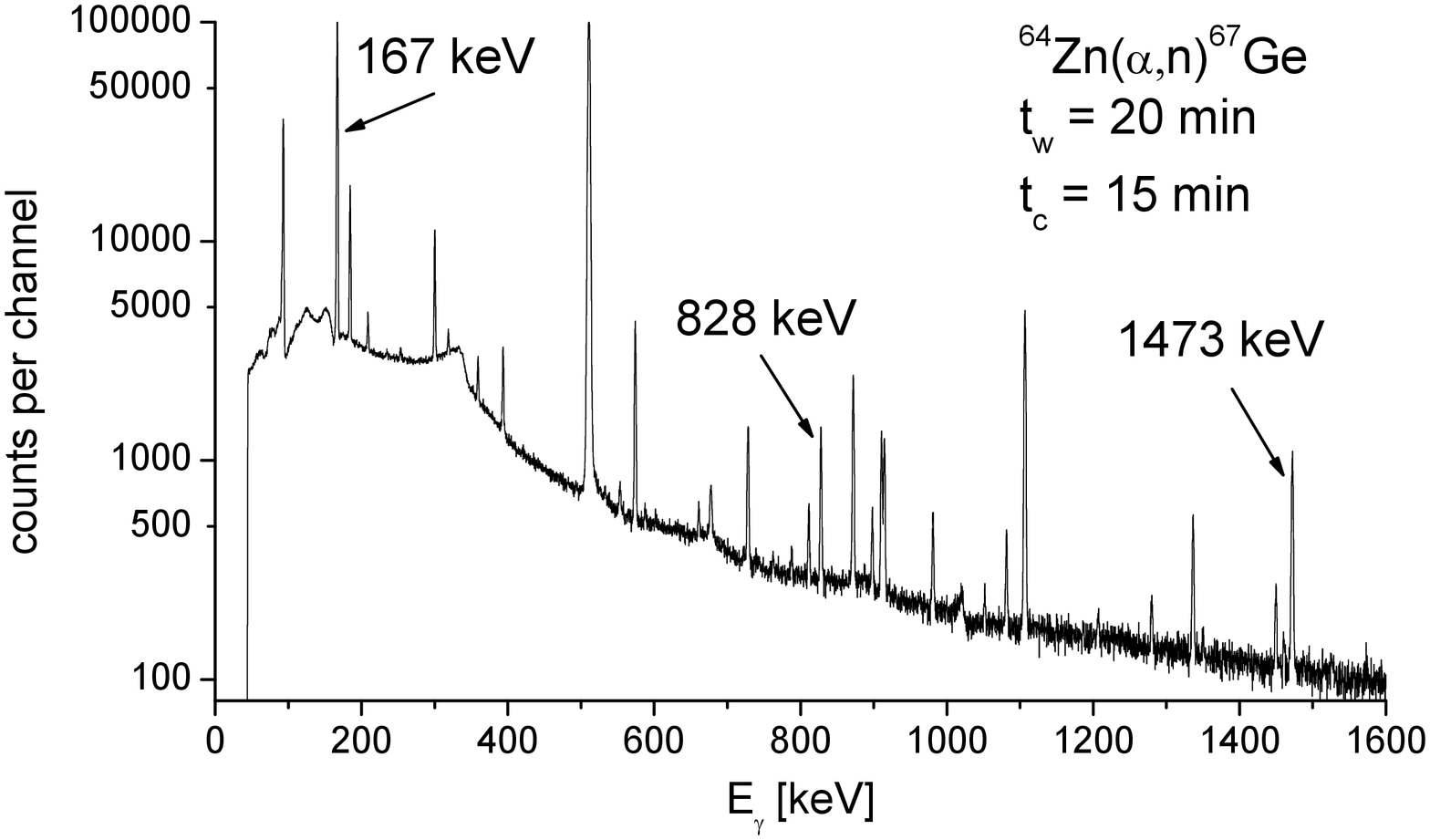} \\
\includegraphics[angle=270,width=\columnwidth]{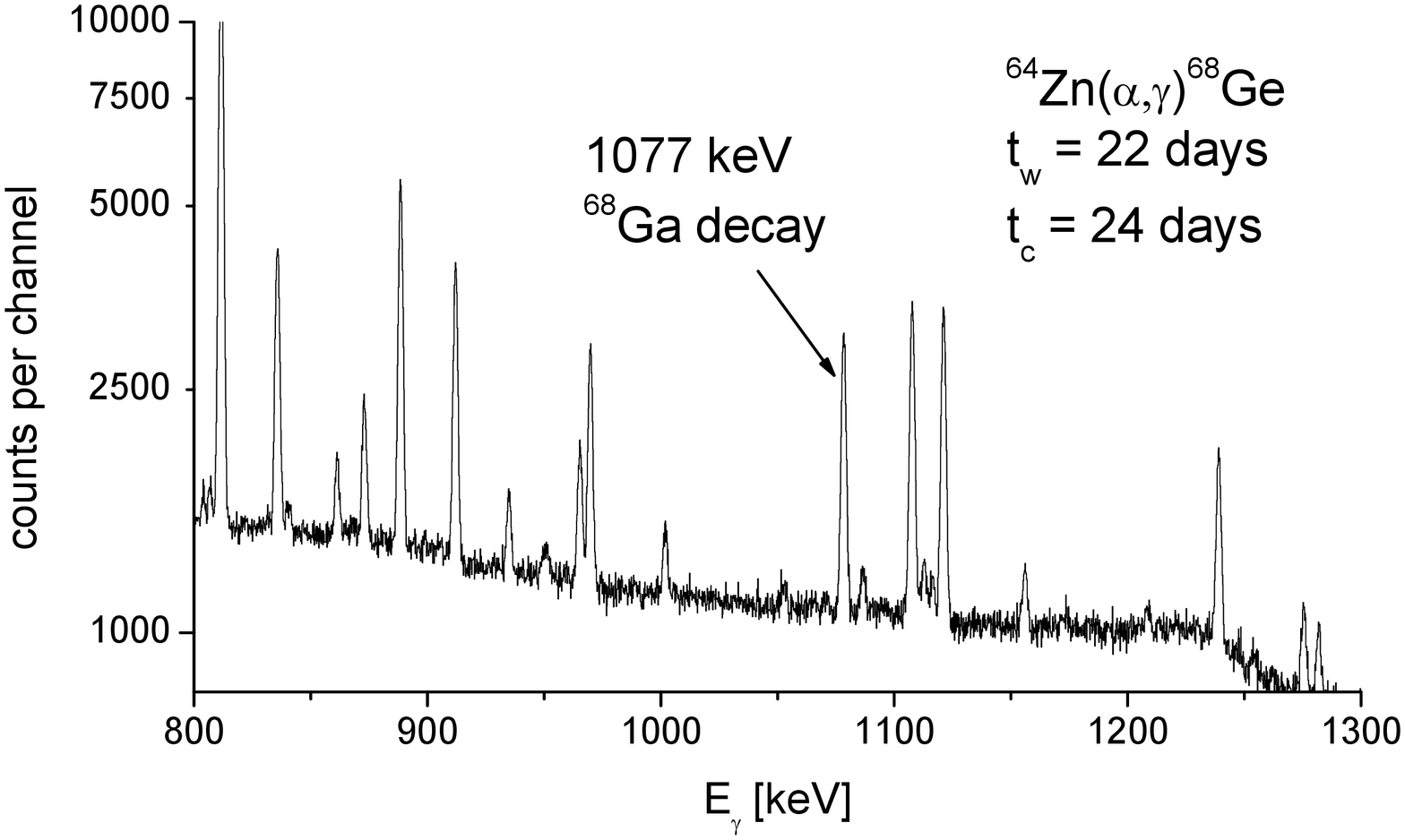}%
\caption{\label{fig:spectra} Gamma-spectra measured on a target irradiated with a 13.2\,MeV $\alpha$-beam. Upper panel: spectrum measured directly after the irradiation where the decay of the \zniv\ran\gevii\ reaction product is dominant. Lower panel: spectrum taken after 22 days of cooling time of the target in order to detect the low activity of the \rag\ reaction product. The peaks used for the analysis are indicated.}
\end{figure}

A typical $\gamma$-spectrum of the \zniv\rap\gavii\ reaction channel was shown in Ref.\,\cite{gyu12}. Here in Fig.\,\ref{fig:spectra} two spectra are shown which were used to determine the \zniv\ran\gevii\ (upper panel) and \zniv\rag\geviii\ (lower panel) cross sections at E$_{c.m.}$ = 12.37\,MeV. The waiting time between the end of the irradiation and the start of counting as well as the durations of the countings are indicated. The peaks used for the analysis are marked by arrows.

Table \ref{tab:results} lists the measured cross sections for the three
reactions. The cross section of the \zniv\rap\gavii\ reaction was measured in
a wide energy range between E$_{\rm c.m.}$ = 5.8 and 12.4\,MeV.
The \zniv\ran\gevii\ reaction was studied from the threshold up to
12.4\,MeV at 9 energies. Owing to the long half-life of the \rag\ reaction
product and the weak $\gamma$-branching of the decay, the activation method is
not optimal for the measurements of the very low \rag\ cross sections.
Therefore, the \rag\ cross section was measured only at a few energies.
Further \rag\ measurements are planned using e.g.\ the AMS method \cite{VERA}. Data for
the three channels \rap , \ran , and \rag\ are available at 12.37\,MeV (shown
in bold face in Table \ref{tab:results}); these data are used for the
determination of the total reaction cross section \stot .

The effective energies were calculated by taking into account the energy loss
of the beam in the target. The uncertainty of the energy is the quadratic sum
of the uncertainty of the beam energy (0.5\,\%) and half of the energy loss in
the target. The uncertainty of the cross section values is the quadratic sum
of the uncertainties from the following components: target thickness (8\,\%),
detector efficiency (5\,\%), current  integration (3\,\%), decay parameters
($<$\,4\,\%), counting statistics (typically about $0.3 - 5$\,\%, in the
worst case $<$\,23\,\%).

\begin{figure}
\includegraphics[angle=270,width=\columnwidth]{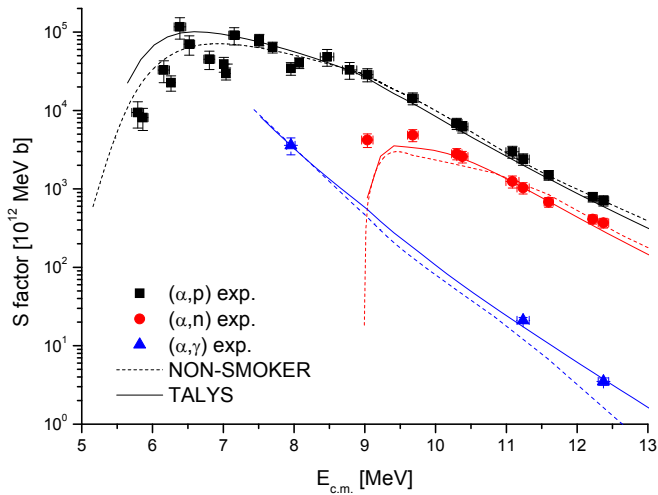}%
\caption{\label{fig:results} (color online) Astrophysical S-factor of the measured reactions and the predictions of statistical model calculations.}
\end{figure}

Figure \ref{fig:results} shows the results of the three reactions in the form
of the astrophysical S-factor. Along with the experimental data the
predictions of two different StM codes, the NON-SMOKER \cite{NS} and TALYS
\cite{TALYS} codes, are plotted using their standard parameters. Both codes
give a good qualitative description of the measured data. 
At $E_{\rm c.m.}$\,=\,12.37\,MeV, where
the total cross section is determined, both codes reproduce well the
\rap\ and \ran\ cross sections. The \rag\ value is well predicted by TALYS and somewhat
underestimated by NON-SMOKER which code is not optimized for high energies. The discussion 
of the low energy cross sections and the astrophysical consequences will be the subject of a 
forthcoming publication.

\begin{table}
\caption{\label{tab:results} Measured cross sections of the three studied
  reactions. See text for details.
}
\begin{ruledtabular}
\begin{footnotesize}
\begin{tabular}{r@{\extracolsep{\fill}}c@{\extracolsep{\fill}}lr@{\extracolsep{\fill}}c@{\extracolsep{\fill}}l@{\extracolsep{2mm}}r@{\extracolsep{\fill}}c@{\extracolsep{\fill}}lr@{\extracolsep{\fill}}c@{\extracolsep{\fill}}l}
\multicolumn{3}{c}{E$^{\rm eff}_{\rm c.m.}$}& \multicolumn{3}{c}{cross section}& \multicolumn{3}{c}{E$^{\rm eff}_{\rm c.m.}$}& \multicolumn{3}{c}{cross section}\\
\multicolumn{3}{c}{[MeV]}& \multicolumn{3}{c}{[mbarn]}& \multicolumn{3}{c}{[MeV]}& \multicolumn{3}{c}{[mbarn]}\\
\hline
\multicolumn{6}{c}{\zniv\rap\gavii~reaction}											&	\multicolumn{6}{c}{\zniv\rap\gavii~reaction~(cont.)}\\
\cline{1-6}\cline{7-12}
5.79	&	$\pm$	&	0.08	&	\multicolumn{3}{c}{(2.65$\pm$0.38)$\cdot10^{-3}$}	&	10.38	&	$\pm$	&	0.08	&	177	&	$\pm$	&	20	\\
5.86	&	$\pm$	&	0.07	&	\multicolumn{3}{c}{(3.03$\pm$0.40)$\cdot10^{-3}$}	&	11.09	&	$\pm$	&	0.09	&	254	&	$\pm$	&	28	\\
6.16	&	$\pm$	&	0.08	&	\multicolumn{3}{c}{(3.73$\pm$0.42)$\cdot10^{-2}$}	&	11.24	&	$\pm$	&	0.09	&	255	&	$\pm$	&	28	\\
6.26	&	$\pm$	&	0.05	&	\multicolumn{3}{c}{(3.69$\pm$0.42)$\cdot10^{-2}$}	&	11.60	&	$\pm$	&	0.06	&	265	&	$\pm$	&	29	\\
6.39	&	$\pm$	&	0.08	&	\multicolumn{3}{c}{(2.94$\pm$0.33)$\cdot10^{-1}$}	&	12.22	&	$\pm$	&	0.06	&	317	&	$\pm$	&	36	\\
6.52	&	$\pm$	&	0.08	&	\multicolumn{3}{c}{(2.76$\pm$0.31)$\cdot10^{-1}$}	&\textbf{	12.37}	&	$\pm$	&	\textbf{0.07}	&	\textbf{344}	&	$\pm$	&	\textbf{40}	\\
6.80	&	$\pm$	&	0.08	&	\multicolumn{3}{c}{(4.38$\pm$0.49)$\cdot10^{-1}$}	&	\multicolumn{6}{c}{\zniv\ran\gevii~reaction}\\
\cline{7-12}
7.01	&	$\pm$	&0.06	&	\multicolumn{3}{c}{(7.08$\pm$0.80)$\cdot10^{-1}$}	&	9.04	&	$\pm$	&	0.08	&	10.6	&	$\pm$	&	1.2	\\
7.04	&	$\pm$	&	0.05	&	\multicolumn{3}{c}{(5.88$\pm$0.66)$\cdot10^{-1}$}	&	9.68	&	$\pm$	&	0.08	&	41.7	&	$\pm$	&	4.6	\\
7.15	&	$\pm$	&	0.08	&	2.52	&	$\pm$	&	0.28	&	10.30	&	$\pm$	&	0.07	&	70.4	&	$\pm$	&	7.7	\\
7.51	&	$\pm$	&	0.04	&	5.63	&	$\pm$	&	0.63	&	10.38	&	$\pm$	&	0.08	&	73.6	&	$\pm$	&	8.1	\\
7.51	&	$\pm$	&	0.04	&	5.95	&	$\pm$	&	0.67	&	11.09	&	$\pm$	&	0.09	&	106	&	$\pm$	&	12	\\
7.70	&	$\pm$	&	0.04	&	7.78	&	$\pm$	&	0.88	&	11.24	&	$\pm$	&	0.09	&	109	&	$\pm$	&	12	\\
7.96	&	$\pm$	&	0.06	&	7.96	&	$\pm$	&	0.89	&	11.60	&	$\pm$	&	0.06	&	120	&	$\pm$	&	13	\\
8.07	&	$\pm$	&	0.05	&	12.4	&	$\pm$	&	1.4	&	12.22	&	$\pm$	&	0.06	&	165	&	$\pm$	&	18	\\
8.47	&	$\pm$	&	0.10	&	36.6	&	$\pm$	&	4.1	&	\textbf{12.37}	&	$\pm$	&	\textbf{0.07}	&	\textbf{179}	&	$\pm$	&	\textbf{19}	\\
8.79	&	$\pm$	&	0.10	&	49.8	&	$\pm$	&	5.6	&	\multicolumn{6}{c}{\zniv\rag\geviii~reaction}		\\
\cline{7-12}
9.04	&	$\pm$	&	0.08	&	72.3	&	$\pm$	&	7.3	&	7.96	&	$\pm$	&	0.06	&	0.92	&	$\pm$	&	0.23	\\
9.68	&	$\pm$	&	0.08	&	123	&	$\pm$	&	13	&	11.24	&	$\pm$	&	0.09	&	2.42	&	$\pm$	&	0.21	\\
10.30	&	$\pm$	&	0.07	&	172	&	$\pm$	&	19	&	\textbf{12.37}	&	$\pm$	&	\textbf{0.07}	&	\textbf{1.81}	&	$\pm$	&	\textbf{0.19}	\\

\end{tabular}
\end{footnotesize}
\end{ruledtabular}
\end{table}

Now we focus on the measurement at the energy $E_{\rm{c.m.}} = 12.37$\,MeV
and determine the sum on the right-hand side ({\it{r.h.s.}}) of
Eq.~(\ref{eq:sum}). The first three terms can be taken directly from our
experimental results in Table \ref{tab:results}. The summed cross section of these three channels is
525\,$\pm$\,57\,mb where for the calculation of the uncertainty the common 
uncertainties of the three measurements have been taken into account. The remaining five terms have
to be estimated using theoretical considerations. Because these terms are
relatively small compared to the dominating \rap\ and \ran\ cross
sections, the sum in Eq.~(\ref{eq:sum}) is mostly defined by our experimental
data.

\raapr : Inelastic scattering is dominated by Coulomb excitation at energies
below the Coulomb barrier. We have performed coupled-channel calculations
with the code ECIS \cite{ECIS} in the
vibrational model using the potential of \cite{McF66} and the deformation
parameter $\beta_2 = 0.242$ which is taken from compilation of Raman {\it et
  al.}\ \cite{Ram01}. 
For the first excited $2^+$ state at $E_x = 992$\,keV 
we find a Coulomb contribution of 34\,mb 
(close to the semi-classical result for Coulomb excitation \cite{Ald56} using
the adopted transition strength of $B(E2) = 20$\,W.u.\ \cite{NDS})
and a smaller nuclear contribution of 12\,mb leading to a
total excitation cross section of 46\,mb with an estimated 20\,\% uncertainty
of 10\,mb. The Coulomb-nuclear interference is practically negligible
at the low energies under study. The cross sections of higher-lying
excited states are much smaller which is confirmed by the spectrum of
inelastically scattered \al\ particles at somewhat higher energies (see e.g.\ 
Fig.~2 of \cite{Scu09}). So we estimate $23 \pm 23$\,mb for the higher-lying
states, i.e.\ a lower limit of zero and an upper limit identical to the
dominating cross section of the first excited $2^+$ state. By summing the
dominating contribution of the first excited $2^+$ state ($46 \pm 10$\,mb) and
the estimated contribution of higher-lying states ($23 \pm 23$\,mb) we find in
total $\sigma$\raapr $= 69 \pm 25$\,mb.

\raaa , \raap , \rapp , \raan : The $Q$-values for all two-particle emission
reactions are strongly negative. 
Thus, the probability to emit one or even two charged
particles from the compound nucleus $^{68}$Ge with very low energies is
extremely small. Calculations 
using the code TALYS \cite{TALYS} with different \al -nucleus optical
potentials always lead to negligible cross sections below 1\,mb for each of
the two-particle emission channels. Thus, we adopt $0.5 \pm 0.5$\,mb for each
of the two-particle channels.

Summing up all the above cross sections, we find a total reaction cross
section of \stot $= 596 \pm 62$\,mb for the {\it{r.h.s.}}\ of
Eq.~(\ref{eq:sum}). Next, this number has to be compared to the total reaction
cross section \stot\ derived from the elastic scattering angular distribution
using Eq.~(\ref{eq:stot}).

Di Pietro {\it et al.}\ \cite{DiP04} have measured \zniv \raa \zniv\ elastic
scattering at the energy $E_{\rm{c.m.}}$\,=\,12.4\,MeV. From their analysis
using Woods-Saxon potentials of volume type
they determine a total reaction cross section of
\stot\ $ = 650 \pm 80$\,mb. A reanalysis of these scattering data using a
folding potential in the real part and an imaginary surface Woods-Saxon
potential
(consistent with a global study of $\alpha$-nucleus potentials at low
energies \cite{Mohr_ADNDT})
leads to a slightly lower value of
610\,mb with a significantly reduced $\chi^2/F$. So we adopt \stot\ $ = 610
\pm 80$\,mb in the following discussion. 
This result also fits nicely into the systematics of so-called reduced cross
sections \sred\ versus reduced energy \ered\ \cite{Far10,Mohr10}
for \zniv \raa \zniv\ scattering data at higher energies
\cite{DiP04,OLDSCAT}; 
however, the
obtained \sred\ for \zniv\ are slightly higher than \sred\ for other 
\al -nucleus systems
\cite{Mohr10,Mohr11}.

The energy difference between the elastic scattering data at 12.4\,MeV and
the activation data at 12.37\,MeV is very small. The calculated \stot\
at 12.37\,MeV differs by less than 0.5\,\% from the
result at 12.4\,MeV. This tiny difference is neglected, and we take
\stot\ $= 610 \pm 80$\,mb for \stot\ in Eq.~(\ref{eq:stot}) at
12.37\,MeV.

The ratio $r = l.h.s./r.h.s$ between the left-hand side and right-hand side in
Eq.~(\ref{eq:sum}) should be unity if the {\it{l.h.s.}} can be taken from
Eq.~(\ref{eq:stot}). The
experimental result of this work $r = 1.02 \pm 0.17$ confirms the validity
of Eqs.~(\ref{eq:stot}) and (\ref{eq:sum}) within about 17\,\% uncertainty
which is the first 
experimental confirmation of Eqs.~(\ref{eq:stot}) and (\ref{eq:sum}) for \al
-induced reactions at low energies around the Coulomb barrier. 
A more precise confirmation requires a reduction of the
uncertainties. Such a 
reduction is possible for the total reaction cross section from elastic
scattering where uncertainties of the order of a few per cent can be achieved
if the angular distribution is measured over the full angular range with small
uncertainties \cite{Mohr10}. Such an experiment is in preparation at ATOMKI. 
It is also planned to measure inelastic scattering cross sections in
this experiment; this will lead to a further reduction of the uncertainty of
$r$ because of a more precise determination of the
\raapr\ cross section in Eq.~(\ref{eq:sum}).

In the previous work \cite{Auce94,Abe94,Ait95} ratios of about $1.1 \le r \le 1.5$
were found for various nuclei at energies above the Coulomb barrier whereas $r
\approx 1$ was found using the latest transmission data \cite{Ing00}. However,
there is no explicit determination of the ratio $r$ in recent literature
\cite{Auce94,Abe94,Ing00} because in
most cases the energies of the transmission data did not exactly match the
energies of the available elastic scattering angular distributions, and thus
also no uncertainty for $r$ is given. The experimental uncertainties for the
transmission data are about $1-3$\,\% in \cite{Auce94} and $4-10$\,\% in
\cite{Ing00}. As can be seen
from Fig.~1 of \cite{Abe94}, different parametrizations of the \al -nucleus
potential 
differ by about $10 - 20$\,\% in the calculation of \stot ; the differences
may result from 
the limited angular range of the angular distributions at higher
energies. Together with an additional uncertainty from the mismatch of the
energies between the transmission data and the scattering data we estimate a
total uncertainty of at least 20\,\% for the ratio $r$ at higher energies.
We recommend to perform a new transmission experiment for a properly chosen
target nucleus and a 
simultaneous study of the elastic and inelastic scattering angular
distributions (with small uncertainties over a broad angular range) at exactly
the same energy. With these data it should be possible to confirm
Eqs.~(\ref{eq:stot}) and (\ref{eq:sum}) also at energies significantly above
the Coulomb barrier with small uncertainties. 

We have also determined the total reaction cross section at the other
energies where we have measured three reaction channels. The summed
cross sections are 366\,$\pm$39\,mb (8.9\,$\pm$1.0\,mb) at 11.24 MeV (7.96 MeV). The
given numbers do not include inelastic cross sections of 62\,$\pm$22\,mb
(16\,$\pm$5\,mb). From these data it is possible to determine \stot\ =
428\,$\pm$44\,mb at 11.24 MeV. We do not provide \stot\ at 7.96\,MeV because
\stot\ is dominated by inelastic scattering and not by our
experimental data. A test of Eq.~(\ref{eq:stot}) is not possible at
these lower energies because experimental angular distributions are
not available. Such scattering experiments will be very difficult in
particular at the lower energy because the elastic scattering cross
section will deviate from the Rutherford cross section by less than
10\,\% here.

In conclusion, this work has confirmed for the first time experimentally that
the total reaction cross section \stot\ of \al -induced reactions from the sum
over all open reaction channels (measured by the activation technique) and
\stot\ from the analysis of elastic scattering angular distributions is
identical. This is an important experimental confirmation of the 
basic quantum-mechanical relations in Eqs.~(\ref{eq:stot}) and (\ref{eq:sum})
which are widely used in the prediction of reaction cross sections e.g.\ in
the statistical model, in particular because some previous studies 
\cite{Auce94,Abe94,Ait95} failed to confirm this identity. A close relation between
reaction cross sections and backward angle elastic scattering was also found by the
analysis of barrier distributions (e.g. \cite{Tim95}). If the validity of
Eqs.~(\ref{eq:stot}) and (\ref{eq:sum}) is assumed {\it{a priori}}, our
result may also be interpreted as consistency check of different methods for
the determination of the total reaction cross section \stot . This result also
strengthens the motivation for the study of elastic \al -scattering at low
energies to determine \stot\ and to obtain a low-energy \al -nucleus potential
for a better prediction of \rag\ capture cross sections
\cite{Mohr_ADNDT,Pal12a}.

\begin{acknowledgments}
We thank Zs.\ Dombradi for the ECIS calculations and A.\ Di Pietro and
V.\ Scuderi for encouraging discussions. This work was supported by ERC St.G. 203175 and OTKA PD104664, K101328 and NN83261
(EuroGENESIS).
\end{acknowledgments}

\end{document}